\documentclass[10pt]{article}
\usepackage{psfig}
\usepackage[cp1251]{inputenc}
\usepackage[russian]{babel}

\begin{document}

\twocolumn[ \noindent{\small\it Astronomy Reports, Vol. 49, No.
12, 2005, pp. 947--957. Translated from
Astronomicheski$\check{\imath}$ Zhurnal, Vol. 82, No. 12, 2005,
pp. 1059--1070. Original Russian Text Copyright \copyright 2005 by
Bajkova.}

\vskip -4mm

\begin{tabular}{llllllllllllllllllllllllllllllllllllllllllllllll}
 & & & & & & & & & & & & & & & & & & & & & & & & & & & & & & & & & & & & & & & \\
\hline \hline
\end{tabular}

\vskip 1.5cm

\centerline{\large\bf A High-Accuracy Method for the Removal of
Point Sources} \centerline{\large\bf from Maps of the Cosmic
Microwave Background}

\bigskip

\centerline{\bf A. T. Bajkova}

\medskip

\centerline{\it Main (Pulkovo) Astronomical Observatory,
St.Petersburg, Russia}

\centerline{\small Received December 28, 2004; in final form, May
18, 2005}

\vskip 1.5cm

{\bf Abstract} --- {\small A new method for removing point radio
sources and other non-Gaussian noise is proposed as a means of
improving the accuracy of estimates of the angular power spectrum
of the cosmic microwave background (CMB). The main idea of the
method is to reconstruct fluctuations of the CMB in places
contaminated by such emission, while traditional methods simply
exclude these regions from consideration, leading to the
appearance of "holes" in the resulting maps. The fundamental
possibility of reconstructing the CMB signal in such holes follows
from the analytical properties of a function with a finite spatial
spectrum (the Silk damping frequency). A two-dimensional median
filter is used to localize the point radio sources.Results of
simulations of the method for maps of modest size are presented.
The efficiency of applying the method to reconstruct the CMB from
data with limited resolution and contaminated by appreciable pixel
noise is investigated. The fundamental possibility of applying the
method to reconstruct the CMB distribution in the region of the
Galaxy is also demonstrated. \copyright {\it 2005 Pleiades
Publishing,Inc.} }

\vskip 1cm
]

\centerline{1.~INTRODUCTION }

\medskip

In connection with numerous experiments on measuring the
anisotropy of the cosmic microwave background (CMB) with the aim
of deriving high-accuracy estimates of the main cosmological
parameters of the Universe [1], it is currently of interest to
find new approaches to data reduction that could increase the
accuracy with which various background contaminating signals of
Galactic and extragalactic origin can be removed from radio
observations of the CMB.This problem has been widely discussed in
the literature [2--4].

\vskip 1.5mm

We will restrict our consideration to the removal of pointlike
radio sources and other non-Gaussian noise, which most strongly
distorts the angular power spectrum of the CMB at high spatial
harmonics (multipoles). This problem has been considered by a
number of authors [5--9].The currently available methods differ
primarily in the means used to localize the contaminating signal,
with the aim of subsequently removing the corresponding sections
from the measured maps.

\vskip 1.5mm

The following three strategies for the removal of pointlike
sources are all natural: (1) reconstructing the point sources,then
subtracting them from the measured CMB maps; (2) excluding
contaminated pixels from the maps; and (3) reconstructing the true
values of the CMB in contaminated pixels.

\vskip 1.5mm

Experience shows that the first strategy leads to larger errors in
the angular power spectrum of the CMB than the other two methods,
since it is not possible to completely accurately reconstruct the
brightness distribution of the point sources when the spatial
spectra of the individual distinguished signals overlap [10]. For
this reason, maps obtained with this method are characterized by
the presence of appreciable non-Gaussian residuals. The second
approach, which requires only the determination of the locations
of contaminated pixels and their elimination before estimating the
CMB power spectrum, is substantially more effective. Precisely
this method is currently the most widely used.

\vskip 1.5mm

However, this second method for the removal of contaminating point
sources is likewise not free of drawbacks. First, the total
effective area of the map that is used to estimate the angular
power spectrum is reduced, leading to appreciable errors for more
distant sections of the CMB. Second, the simple elimination of
contaminated pixels does not fully restore the Gaussian statistics
of the CMB, since it leaves "holes" in the maps that are used. In
addition, the sharp boundaries of the cutout sections lead to
so-called Gibbs phenomena in the behavior of the spectrum, which
are especially strongly manifest at high spatial harmonics,and can
appreciably complicate studies of the secondary anisotropy of the
CMB.

\vskip 1.5mm

Note as well that increasing the resolving power of a system used
to measure the CMB anisotropy will lead to the detection of a
larger number of weak sources [11], so that an increasingly large
number of pixels in the CMB map are contaminated by these signals.
A special problem is also presented by the strong contamination
due to Galactic emission. Simple removal of the zone of Galactic
emission, which comprises an appreciable fraction of the overall
area of the celestial sky, unavoidably leads to appreciable errors
when estimating the angular power spectrum compared to the use of
completely uncontaminated data.

\vskip 1.5mm

Obviously, if the third strategy is realizable in practice, which
depends primarily on the noise characteristics of the apparatus,
it will be free of the backs of the first two methods.Therefore,we
propose and develop here a new method for the removal of
high-multipole non-Gaussian noise, including both the localization
of such noise and reconstruction of the true values of the CMB
anisotropy in the contaminated regions. For the specific
conditions considered in this paper, this could lead to an
increase in the accuracy with which the angular spectrum of the
CMB can be constructed,compared to the simple elimination of
contaminated map pixels.

\vskip 1.5mm

Thus,our aim is to present and develop a new method for
eliminating the contamination of a CMB signal by point sources and
other sources of high-multipole noise. We estimate the internal
accuracy of the method and its stability to the input noise, and
also investigate its application to data obtained with finite
resolution for various levels of the instrumental pixel noise,
thereby evaluating possibilities for applying the method to real
systems.

\vskip 1.5mm

The following sections of the paper present a statistical
description of the CMB anisotropy; describe our model for the
observed map, the method used to localize the point sources, and
the method used to reconstruct the CMB values in
noise-contaminated regions; and present the results of numerical
simulations carried out for maps with modest angular dimensions
but having the same statistical properties as the CMB distribution
over the entire celestial sphere.

\bigskip

\centerline{2.~MODEL FOR AN OBSERVED MAP} \centerline{OF THE CMB }

\medskip

The distribution of the CMB temperature over the celestial sphere
can be presented by the following expansion in spherical harmonics
$Y_{\ell}^m(\theta,\phi)$ [12]:

$$
\frac{\Delta T}{T}=\sum_{\ell,m}a_{\ell m}
Y_{\ell}^m(\theta,\phi),
$$
where $T$ and $\Delta T$ are the mean temperature and temperature
fluctuations of the CMB and the $a_{\ell m}$ are the coefficients
of the expansion.  The angular power spectrum of the fluctuations
$C_{\ell}$ is determined as the mean square of the coefficients
$a_{\ell m}$:

$$
C_{\ell}=<|a_{\ell m}|^2>,
$$
where $\ell$ is the multipole number.

\vskip 1.5mm

If the fluctuations in the early Universe satisfy Gaussian
statistics, as is expected in most cosmological theories, each
coefficient $a_{\ell m}$ should be statistically independent.
Thus, the power spectrum $C_{\ell}$ provides a full statistical
description of the CMB anisotropy, which is a fundamental
characteristic of the Universe that can be obtained directly from
observations via a spherical-harmonic analysis.

\vskip 1.5mm

For Gaussian fields,the expansion coefficients $a_{\ell m},
\ell\ne 0,$, likewise represent Gaussian fields with random
phases, zero means,and the dispersions
$$
<a_{\ell m} a_{\ell^{'}m^{'}}>=\delta_{\ell
\ell^{'}}\delta_{mm^{'}} C_\ell,~~ \ell\ne 0.
$$

\vskip 1.5mm

Here, we consider sections of the sky that are modest in size. In
this case, it is expedient to operate with the Gaussian fields in
a flat, two-dimensional space, and to replace the
spherical-harmonic analysis with a Fourier analysis, which
appreciably simplifies the testing of proposed methods. The CMB
fluctuations $\Delta T=T-<T>$ can then be generated via the simple
calculation of the Fourier series [12]

\begin{equation}
\frac{\Delta
T(\theta_x,\theta_y)}{T}=\sum_{n_u=0}^{N_u-1}\sum_{n_v=0}^{N_v-1}D(n_u,n_v)
\end{equation}

$$
\times\exp[i\frac{2\pi}{L}(n_u\theta_x+n_v\theta_y)],
$$
where $L$  is the linear size of the region considered in radians,
$(\theta_x, \theta_y)$ are Cartesian coordinates on the sky (in
the spatial region), and $(n_u,n_v)$ is the number of the Fourier
component $D(n_u,n_v)$ in the region with spatial frequencies $u$
and $v$.

\vskip 1.5mm

The amplitudes of the Fourier components $D(n_u,n_v)$ obey a
Gaussian distribution with zero mean and the dispersion
\begin{equation}
<|D(n_u,n_v)|^2>=C_{\ell},~~~\ell=\frac{2\pi}{L}\sqrt{n_u^2+n_v^2},
\end{equation}
while the phases are uniformly distributed in the interval $(0,
2\pi)$. Here, $C_l$ is the angular power spectrum of the CMB
temperature when it is expanded in spherical harmonics.

\vskip 1.5mm

Relation (2) describes the circular symmetry of the power
spectrum, i.e., its independence of the azimuthal number $m$,
which we will use below as additional {\it a priori} information
when reconstructing the CMB fluctuations in noise-contaminated
regions.

\vskip 1.5mm

Another, more important, characteristic of the power spectrum is
its finite spatial extent,which follows from the existence of the
so-called Silk damping frequency ($\ell_D$)[13], above which the
power-spectrum fluctuations fall off sharply and the contribution
of the CMB to the total observed signal becomes negligibly small
with increasing frequency. The finiteness of the spectrum enables
us to apply the theory of analytical functions [14] to describe
the CMB fluctuations, which implies the possibility of
reconstructing functions over an entire region based on knowledge
for part of the region or at some set of points [15]. This
principle lies at the basis of the proposed algorithm for
reconstructing the CMB fluctuations in regions contaminated by
noise.

\vskip 1.5mm

When modeling the background of point sources (PS),we assume that
they are randomly distributed over the sky in accordance with a
Poisson law.The instrumental pixel noise is "white" Gaussian noise
with zero mean.The measured CMB map can be represented by the
model
\begin{equation}
CMB_{meas} = (CMB + PS)\ast BEAM + N,
\end{equation}
where $N$ represents noise, $BEAM$ the antenna beam, and $\ast$ a
linear convolution.

\vskip 1.5mm

The required data processing consists of solving an equation of
the form of the convolution (3) for the CMB signal. The sequence
of operations we have used during the reconstruction of the CMB is
the following.

\vskip 1.5mm

(1)The noise is filtered using a Wiener filter or a modification
of such a filter that does not distort the form of the angular
power spectrum [16]; this yields an estimate of the signal
$(CMB+PS)\ast BEAM$.

\vskip 1.5mm

(2)This quantity is deconvolved from the antenna beam using a
regularized inversion filter [17,18], in order to derive an
estimate of the signal $CMB+PS$.

\vskip 1.5mm

(3)Positions contaminated by point sources are localized using a
median filter (see Section 4).

\vskip 1.5mm

(4)The CMB signal is reconstructed at the contaminated positions
using the method proposed in the following section.

\bigskip

\centerline{3.~RECONSTRUCTION} \centerline{OF THE CMB
FLUCTUATIONS} \centerline{AT CONTAMINATED POSITIONS}

\medskip

The proposed method for reconstructing the fluctuations in holes
in a CMB map is a modification of the algorithm of Fienup [19],
intended for the reconstruction of an image of an object with a
finite carrier from the amplitude of its Fourier spectra (the
phase problem). In our modified version, we determine the
limitation on the spectral region using information about the
finiteness and circular symmetry (2) of the amplitude
spectrum,rather than the amplitude spectrum itself, and determine
the limitation on the spatial region using the known map values.

\vskip 1.5mm

The algorithm is iterative, and consists of the following sequence
of operations.

\vskip 1.5mm

(1)~An initial approximation for the map is made. We recommend the
use of the initial map with zero brightness in specified locations
(holes) as a first approximation.

\vskip 1.5mm

(2)~The Fourier transform of the initial approximation is
calculated, bringing about a transformation to the
spatial-frequency domain.

\vskip 1.5mm

(3)~A condition for the spatial limitation of the fluctuation
spectra,which comes about because the values of the Fourier
components derived in the first step with numbers $\ell>\ell_D$
have been set equal to zero,is imposed.An additional constraint on
the region of spatial frequencies that appreciably speeds up the
convergence of the algorithm is the circular symmetry of the power
spectrum of the fluctuations. To satisfy this last condition, the
form of the Fourier components is modified so that the power
spectrum has a form that is consistent with relation (2). For each
value of $\ell$, the squared amplitudes of the spectra
measurements are averaged in radius over a length $\ell$, after
which these measurements are replaced by values with the derived
squared amplitudes, retaining the phases.

\vskip 1.5mm

(4)~The inverse Fourier transform of the spectrum obtained in the
previous step is taken, bringing about a transformation to the
domain of the CMB map.

\vskip 1.5mm

(5)~The constraints on the spatial region of the map are imposed.
The brightness values outside the holes in the map are replaced by
the known values, while the values inside holes are not changed.

\vskip 1.5mm

(6)~The Fourier transform of the map obtained in step 5 is taken.

\vskip 1.5mm

(7)~Return to step 3 until the image obtained in step 5 ceases to
change in accordance with a chosen convergence criterion.

\vskip 1.5mm

Simulations show (see Section 6) that this modified Fienup
algorithm leads rather rapidly to the correct solution. For this
to be the case, the number of unknown values of the CMB
fluctuations should be approximately half the number of knowns, if
the discretization frequency of the map is twice its upper spatial
frequency. If the discretization frequency is increased, a larger
number of map values can be reconstructed, since the insufficient
information in the spatial domain can be compensated by
information in the spatial-frequency domain.

\vskip 1.5mm

Obviously, a high computational speed for the method can be
reached via the application of Fast Fourier Transform algorithms.

\bigskip

\centerline{4.~METHOD FOR LOCATING} \centerline{CONTAMINATED
SECTIONS}

\medskip

The proposed method for localizing the point sources is based on
applying a two-dimensional median filtration,which is a row-by-row
column algorithm consisting of one-dimensional $n$-point median
filtrations [20,21]. The number $n$ is taken to be odd. If
$n=2k+1$, one-dimensional median filtration consists of ascribing
to the current value for the sequence the mean of the series that
is obtained when the $(2k +1)$-point sequence is placed in
increasing order, with the first $k$ values located to the left
and the last $k$ values to the right of the given value. As a
result of applying this operation to a one-dimensional series, all
impulsive noise is eliminated. In our case, such noise corresponds
to point sources.

\vskip 1.5mm

To obtain the series of removed sources itself, we subtract the
output map from the input map. Obviously, since the nonlinear
transform is applied to the useful signal together with the noise,
the resulting difference map obtained from the subtraction
contains additional noise, whose magnitude is appreciably lower
than the desired impulsive noise when $n$ is chosen correctly.
Sections of the resulting difference map that are contaminated by
point sources can be identified by applying a cutoff to the levels
of this map. The lower the cutoff level, the more pixels are
subject to distortion. In our case, the cutoff level was usually
chosen to be fairly low (10\% of the peak value of the CMB
fluctuations), in order to avoid missing genuine places that are
contaminated by weak radio sources. On the other hand, the number
of identified points must not be too large to ensure convergence
of the subsequent reconstruction of the CMB-fluctuation values at
these points.

\vskip 1.5mm

In contrast to numerous methods that have been proposed in the
literature (wavelet analysis [7,8], clean and maximum-entropy
deconvolution [10], optimal linear filtration [9], etc.), the
proposed method for localizing the point sources is distinguished
by its simplicity and the speed with which it operates, which is
important in the reduction of large datasets obtained over the
entire celestial sphere.

\vskip 1.5mm

Obviously, this method will operate more reliably the brighter the
point sources and the fewer the number of point sources in the map
being analyzed. We have investigated the method for the cases of
both relatively bright (resolved) and relatively weak (unresolved)
radio sources, whose number grows with the resolution of the
instrument used [11].

\bigskip

\centerline{5.~FILTRATION} \centerline{OF THE INSTRUMENTAL NOISE}
\centerline{AND DECONVOLUTION}

\medskip

We used a modification of a Wiener filter that preserves the form
of the angular power spectrum [16], called a power filter, for the
filtration of the instrumental white noise. A power filter can be
applied in two ways. In the first, the dispersion of the
instrumental (pixel) white noise is taken to be known {\it a
priori}, whereas in the second, it is not. We present a
preliminary estimate of this noise based on applying a median
filter to the input signal. The dispersion of the instrumental
noise was determined by subtracting the output signal of the
median filter from the measured map, $CMB_{meas}$. We present here
results obtained using the second approach. Reconstruction of the
map taking into account the antenna beam was carried out using a
regularized inversion filter containing a regularizing parameter
that depends on the level of the residuals after filtration of the
noise [17,18].

\bigskip

\centerline{6.~SIMULATION RESULTS}

\medskip

In this section, we present the results of applying the proposed
method to model CMB measurements. The anisotropy of the CMB was
simulated using a function for the angular power spectrum that
corresponds to a standard  $\Lambda$CDM cosmological model for the
Universe with $\Omega_b h^2$=0.02, $\Omega_{\Lambda}$=0.65,
$\Omega_m$=0.3, $h$=0.65, and $n$=1. We numerically generated a
map of the CMB $7.5^o\times 7.5^o$ in size in accordance with
formulas (1)and (2); this map, therefore, possesses all the
properties of the CMB fluctuations distributed over the entire
celestial sphere. The upper frequency of the angular power
spectrum of the CMB corresponds to the multipole $\ell$=1536.

\medskip

\centerline{\it 6.1.~Effect on the CMB Reconstruction}
\centerline{\it in Point-Source Contaminated Locations}
\centerline{\it in Systems with High Angular Resolution}

\medskip

The goal of this experiment (experiment 1) is to illustrate the
proposed method and to estimate the effect of applying the method
in systems with high angular resolution that detect a large number
of relatively weak radio sources.We tested the method assuming
that the preliminary noise filtration and deconvolution were
already carried out (see the end of Section 2). This was done with
the aim of estimating the internal accuracy of specifically the
method for reconstructing the CMB in holes in the initial CMB map.

\vskip 1.5mm

We took the map to be 64$\times$64 pixels in size, which
corresponds to discretization at the Nyquist frequency for the
specified maximum multipole of the CMB spectrum,$\ell_{max}$=1536.
The results of this numerical experiment are shown in Fig.1, which
presents (a) the initial (model)map of the CMB fluctuations; (b)
the map of the CMB contaminated by radio sources ($CMB +PS$; 234
weak sources with amplitudes exceeding the CMB-fluctuation peak by
a factor of three, distributed randomly over the map according to
a Poisson distribution); (c) the map obtained by processing the
map in (b) with a row-column, three-point median filter; (d) the
map obtained by replacing the CMB values at the detected
contaminated points with zero brightness; (e) identified sections
of contamination (which, in general, differ from the intrinsic
distribution of point sources, see Section 3)obtained by applying
an amplitude cutoff of 10\% of the peak CMB fluctuation to the
difference of maps (b) and (c); and (f) the map of the CMB
reconstructed using the modified Fienup algorithm,which nearly
precisely coincides with the input model CMB map in (a).

\vskip 1.5mm

Figure 2 presents the angular power spectra of the maps; the solid
curve refers to the original and reconstructed CMB maps, the
dotted curve to the contaminated map (b), the short-dashed curve
to the map obtained via median filtration (c), and the long-dashed
curve to the map with zero brightness in contaminated locations
(d). The dispersions of the signals represent a quantitative
characteristic of the experiment,and are presented in Table 1. We
can estimate the effect of applying our approach compared to the
traditional approach that does not reconstruct the CMB
fluctuations in holes by comparing the angular power spectra in
Fig.2 (as well as the dispersions of the corresponding residual
maps).

\vskip 1.5mm

We can see that simple removal of 234 contaminated measurements
from the CMB map leads to distortion of the true angular power
spectrum, especially at high harmonics: this distortion is from
30\% to 100\% when $\ell$>1000. Applying the proposed method to
remove the point sources leads to virtually no such distortion.
Simulation of the case of a background of unresolved point sources
with amplitudes a factor of three smaller than the maximum CMB
fluctuations likewise yielded very high accuracy for the
reconstruction in the absence of pixel noise.

\vskip 1.5mm

To investigate the stability of the method with regard to the
input instrumental noise,we added Gaussian white noise with a
signal-to-noise ratio of SNR$\approx$10 to the map in panel (b),
taking the signal level to be that of the CMB fluctuations. The
dash-dotted curve in Fig.2 depicts the angular power spectrum of
the reconstructed map in this case. We can see that small
variations in the input data gave rise to only small changes in
the solution; i.e., the proposed algorithm displays a fairly high
stability. Analysis of these results indicates that simple zeroing
of the CMB values, even in precisely determined contaminated
locations, yields appreciable errors in the reconstruction of the
angular power spectrum, especially on small angular scales. If the
contaminated sections occupy a large area (Fig.1e), the error in
the desired angular power spectrum can be still larger, possibly
leading to substantial errors in derived estimates of cosmological
parameters.

\vskip 1.5mm

Thus, in this experiment, without allowance for the real
resolution of a system and the input instrumental noise, it was
possible to reach very high accuracy in the CMB-map
reconstruction. As we indicated above, this testifies to a high
internal accuracy of the proposed method for reconstructing the
CMB in holes in a map. Note that convergence of the algorithm to
the required solution can also be achieved without placing
constraints on the circular symmetry of the power-spectrum
fluctuations.However,use of this information as an additional
constraint appreciably speeds up the process of convergence.
Experience shows that,when the map is discretized with a frequency
that is twice the bandwidth of the spectrum signal, high accuracy
in the reconstruction is achieved only if the number of unknowns
is no more than one third of the total number of pixels in the
map. As is shown above,the method displays a fairly high degree of
stability to the noise in the input data; i.e., small variations
in these data correspond to only small variations in the
reconstruction. When the input noise level is high, a preliminary
filtration should be applied using a power filter (see Section 5),
as was done in the experiments discussed in the following section.

\medskip

\centerline{\it 6.2.~Effect on the CMB Reconstruction }
\centerline{\it in Contaminated Locations in the Model}
\centerline{\it CMB$_{meas}$=(CMB + PS )$\ast $BEAM + N}

\medskip

Let us make the problem more complex and consider a model for the
observed CMB map that corresponds to expression (3), which
includes the effect of the limited resolution of the system and
the input pixel noise.In the following set of experiments,the map
size was 128$\times $128 pixels, the pixel size was
3.51$^{\prime}$, and the full width at half maximum of the antenna
beam 10.53$^{\prime}$. For an antenna with a diameter of 1 m, this
resolution corresponds to an observing frequency of 98 GHz.These
system parameters are close to those for one of the channels of
the PLANCK mission [11].

\vskip 1.5mm

We carried out four experiments based on these parameters
(experiment group 2), with various levels of white pixel
noise,beginning with zero noise. The dispersions of the
input,noise,reconstructed,and residual maps can be used as a
measure of the accuracy of the reconstruction of the CMB
anisotropy, and are given in Table 2. The results are shown in
Fig.3, where the numbers (columns)indicate the experiment number
and the letters (rows) indicate the physical meaning of the map:
(a) measured CMB map satisfying (3);( b) reconstructed CMB map
with holes in contaminated pixels, including the effect of the
antenna beam; (c) residual map, equal to the difference of the
given CMB map and the reconstructed map shown in row "b"; (d)
reconstructed CMB map obtained using our method to interpolate
functions in the holes; (e) residual map,equal to the difference
of the given CMB map and the reconstructed map shown in row "d".
Obviously, each column of maps presented in the figure corresponds
to a particular value for the pixel-noise level.

\vskip 1.5mm

The results presented in Fig.3 were obtained using pixel-noise
filtration based on applying a median filter to derive a
preliminary estimate of the noise dispersion (see Section 5). The
effect of the antenna beam in the reconstructed map was included
using a regularized inversion filter.

\vskip 1.5mm

Analysis of the results presented in Fig.3 shows that the main
advantage of the proposed method for removing point sources is
that the residual maps shown in row "e" are free from non-Gaussian
features, and resemble residual Gaussian noise, whose amplitude
depends on the level of the input Gaussian  noise (see the
dispersion in Table 2). As we can see in map row "c", the
traditional method, which simply excludes contaminated pixels from
the CMB map, does not fully eliminate the contribution of the
point sources, which are manifest as small non-Gaussian features
determined by the magnitude of the CMB fluctuations in the
contaminated locations. In addition, the sharp edges of the cutout
sections lead to high-frequency noise that extends far beyond the
frequency edge,which corresponds in our case to $\ell_{max}$=1536.

\vskip 1.5mm

We illustrate the effect of applying our method in pure form in
Fig.4, which shows (a) the angular power spectra of the convolved
map $CMB\ast BEAM$ without (solid curve) and with holes, whose
size is determined by the size of the antenna beam (dashed curve),
and (b) the same spectra for the input CMB map. We can see from
the curves that the presence of holes in the maps leads to errors
of about 13\%i n the angular power spectrum of the CMB for
multipoles $\ell$=200--300 and errors of about 30\% for multipoles
$\ell$=450--550. The relative errors are even higher for higher
$\ell$ values.

\vskip 1.5mm

Note that, in spite of the fact that the radio sources considered
are point sources (i.e.,they are essentially $\delta$ functions),
we have taken the contaminated locations to include a region
surrounding the source coordinates,determined by the
characteristics of the inverse filter,which does not provide an
ideal reconstruction of a $\delta$ function. The presence of even
a small amount of input noise requires regularization of the
algorithm, which leads to a solution with finite resolution. To
ensure reliable exclusion of all contaminated pixels,we have
assumed that the region distorted by each point source occupies an
area equal to the area of the base of the antenna beam, which
exceeds somewhat the total area of the excluded CMB measurements.
However, this is not a serious concern for our method,since we
reconstruct the intrinsic CMB values in the resulting holes.

\vskip 1.5mm

Obviously, the maximum effect from reconstructing the CMB signal
in contaminated locations is achieved in the absence of pixel
noise. The quantitative effect is lowered as the pixel-noise level
is increased, although the qualitative effect of a complete
elimination of non-Gaussian features from the CMB maps is
maintained even in the presence of fairly high noise levels.

\vskip 1.5mm

Figures 5a and 5b present the angular power spectra of the signals
corresponding to experiments 2.1 and 2.4, respectively.The results
obtained in experiments 2.2 and 2.3 occupy intermediate positions
and are not shown. The solid curve shows the angular power
spectrum of the initial CMB map,the dotted curve the measured CMB
map, which satisfies (3), the dashed curve the angular spectrum of
the reconstructed CMB map with the pixels distorted by point
sources zeroed,and the dash-dotted curve the CMB map with
reconstructed values for the fluctuations at the distorted
locations.

\vskip 1.5mm

We can see from Fig.5a that it is possible to reconstruct the CMB
map nearly perfectly in the absence of pixel noise. The error in
the reconstruction is determined only by the error in the
inversion filtration. The effect of applying our method in this
case is close to that shown in Fig.4.

\vskip 1.5mm

Figure 5b shows that a high level of pixel noise leads to a
substantial loss in the accuracy of the reconstructed CMB angular
power spectrum,for both a simple exclusion of contaminated
sections and the reconstruction of the CMB signal in
holes.However, the accuracy is nonetheless higher in the latter
case --- quite appreciably for multipoles $\ell$=24--500. The
dispersion of the residual-noise map for the specified interval of
the input pixel noise is 25--27\% lower due to the full removal of
point sources from the CMB maps (Table 2).

\vskip 1.5mm

Although the simulations show that the effect of applying our
method decreases with growth in the pixel-noise level in a natural
way, it nonetheless remains fairly high for a system with noise
characteristics similar to those planned for the PLANCK mission
[11].

\vskip 1.5mm

The method for reconstructing the CMB signal proposed here can
also be applied in connection with constructing catalogs of point
sources. For this, it is sufficient to subtract the reconstructed
CMB map from the estimated $CMB+PS$ map.

\medskip

\centerline{\it 6.3.~Reconstruction of the CMB Anisotropy}
\centerline{\it over a Wide Region of the Map}

\medskip

From the point of view of enhancing the accuracy of an estimated
angular power spectrum based on CMB measurements over the entire
celestial sphere, it is of interest to consider the reconstruction
of the CMB signal in the zone of the Galaxy where powerful
non-Gaussian noise is observed.

\vskip 1.5mm

In the traditional strategy, this region of the sky is simply not
taken into account, which clearly leads to a loss of accuracy in
the derived CMB angular power spectrum compared to that which
would be obtained using a full set of undistorted data. Let us
consider the reconstruction of the CMB signal in the zone of the
Galaxy using the same method as in the previous experiment. Let
the noise occupy the middle part of the map in the form of a band
that is elongated in the horizontal direction. We cut out the
contaminated region from the map and reconstruct the absent
components of the CMB in this region using our method (experiment
3).

\vskip 1.5mm

Our simulations show that it is possible to obtain a nearly
perfect reconstruction of the CMB, right up to the case when the
width of the contaminated band comprises a third of the linear
size of the map. Further increase in the width of the band leads
to a growth in the errors of the reconstruction. Simulation
results for band widths comprising about 30\% of the linear size
of the map are presented in Fig.6, which shows the (a) initial CMB
map, (b) CMB map with the Galactic band cut out, and (c)
reconstructed CMB map. The angular power spectra are presented in
Fig.7, where the solid curve corresponds to the initial CMB map,
the dashed curve to the map with the band cut out, and the
dash-dotted curve to the reconstructed map.

\vskip 1.5mm

Analysis of the maps presented in Fig.7 shows that simple
elimination of sections of the CMB occupying a substantial area
leads to appreciable errors in the derived angular power spectrum,
which can reach 50\% (for example, for multipoles
$\ell$=200--300). Applying our method led to a virtually perfect
reconstruction.

\vskip 1.5mm

Thus, this experiment provides hope that reconstructing the CMB in
the zone of the Galaxy based  on real measurements over the entire
celestial sphere can substantially improve the accuracy with which
the real angular power spectrum of the CMB can be derived.

\bigskip

\centerline{7.~CONCLUSION}

\medskip

We have investigated the fundamental possibility of improving the
accuracy of estimates of the angular power spectrum of the CMB by
reconstructing the CMB anisotropy in regions contaminated by point
radio sources and other high-multipole noise. A series of examples
have been used to demonstrate the effect of applying this
approach, compared to the currently standard strategy of simply
excluding contaminated sections of CMB maps.

\vskip 1.5mm

We have shown that, in the absence of instrumental noise,the
proposed method ensures the removal of point sources and other
non-Gaussian multipole noise with very high accuracy,and is fairly
stable to variations in the noise level in the input data.

\vskip 1.5mm

The main advantage of the method is that it is able to fully
remove non-Gaussian features from CMB maps, while the simple
exclusion of contaminated locations does not achieve this goal.
The full removal of residual point sources from the maps makes it
possible to appreciably lower the dispersion of the residual
noise, even in the presence of a relatively high level of input
pixel noise. Our simulations show that the effect of applying our
method decreases in a natural way as the input noise level grows,
but nonetheless remains fairly high for noise characteristics
corresponding to those planned for the PLANCK mission.

\vskip 1.5mm

We have also shown the fundamental possibility of enhancing the
accuracy of estimates of the angular power spectrum of the CMB by
reconstructing the CMB anisotropy in the zone of the Galaxy where
the strongest level of background contamination is observed.

\bigskip

\centerline{ACKNOWLEDGMENTS}

\medskip

This work was partially supported by the Basic Research Program of
the Presidium of the Russian Academy of Sciences "Nonstationary
Phenomena in Astronomy".

\bigskip

\centerline{REFERENCES}

\medskip

\noindent 1.~E.Gawiser and J.Silk, Phys. Rep. {\bf 333}, 245
(2000).

{\small
\noindent 2.~M.Tegmark and G.Efstathiou, Mon. Not. R.
Astron. Soc. {\bf 281}, 1297 (1996).

\noindent 3.~M.Tegmark, A.de Oliveira-Costa, and A.Hamilton, Phys.
Rev. D {\bf 68},123523 (2003).

\noindent 4.~V.Stolyarov, M.P.Hobson, M.A.J.Ashdown, {\it et al.},
Mon. Not. R. Astron. Soc. {\bf 336}, 97 (2002).

\noindent 5.~M.Tegmark and A.de Oliveira-Costa, Astrophys.J. {\bf
500}, L83 (1998).

\noindent 6.~M.~P.~Hobson, R.~B.~Barreiro, L.~Toffolatti, {\it et
al.}, Mon. Not. R. Astron. Soc. {\bf 306}, 232 (1999).

\noindent 7.~J.~L.~Sanz,~R.~B.~Barreiro,~L.~Cayon, {\it et al.},
Astron. Astrophys. {\bf 140}, 99 (1999).

\noindent 8.~P.~Vielva, E.~Martinez-Gonzalez, L.~Cayon, {\it et
al.}, Mon. Not. R. Astron. Soc. {\bf 326}, 181 (2001).

\noindent 9.~R.~Vio, L.~Tenorio, and W.~Wamsteker,
Astron.Astrophys. {\bf 391}, 789 (2002).

\noindent 10.~A.~T.~Bajkova, Izv. Vyssh. Uchebn. Zaved., Radiofiz.
{\bf 45}, 909 (2002); Radiophys. Quantum Electron. {\bf 45}, 835
(2002).

\noindent 11.~M.~Bersanelli {\it et al.}, {\it COBRAS/SAMBA, ESA
Report D/SCI} (1996).

\noindent 12.~J.~R.~Bond and G.~Efstathiou, Mon. Not. R. Astron.
Soc. {\bf 281}, 655 (1987).

\noindent 13.~J.Silk, Nature {\bf 215}, 1155 (1967).

\noindent 14.~Ya.~I.~Khurgin and V.~P.~Yakovlev, {\it Finite
Functions in Physics and Technology} (Nauka, Moscow, 1971) [in
Russian].

\noindent 15.~G.~I.~Vasilenko and A.~M.~Taratorin, {\it Image
Reconstruction} (Radio i Svyaz', Moscow, 1986)[in Russian].

\noindent 16.~K.M.Gorski, astro-ph/9701191 (1997).

\noindent 17.~A.~N.~Tikhonov and V.~Ya.~Arsenin, {\it Solutions of
Ill-Posed Problems} (Halsted, New York, 1977; Nauka, Moscow,
1986).

\noindent 18.~A.~V.~Goncharskii, A.~M.~Cherepachshuk, and
A.~G.~Yagola, {\it Numerical Method for Solving Inverse Problems
in Astrophysics} (Nauka, Moscow, 1978)[in Russian].

\noindent 19.~J.~R.~Fienup, Opt. Lett. {\bf 3}, 27 (1978).

\noindent 20.~B.~Yustusson, in {\it Two-Dimensional Digital Signal
Processing II.Transforms and Median Filters}, Ed. by T.~S.~Huang
(Springer-Verlag, New York, 1984; Radio i Svyaz', Moscow, 1984),
p. 156.

\noindent 21.~S.~G.~Tyan, in {\it Two-Dimensional Digital Signal
Processing II. Transforms and Median Filters}, Ed. by T.~S.~Huang
(Springer-Verlag, New York, 1984; Radio i Svyaz', Moscow, 1984),
p. 191.

}

\bigskip

{\it Translated by D.Gabuzda}

\clearpage \onecolumn

\newpage

\begin{center}
\centerline{{\bf Table 1.}~Dispersion $\sigma$ (in units of
$10^{-5}$) of the maps in experiment 1} \vskip 2.5mm
\begin{tabular}{|c|c|c|c|c|c|c|c|c|} \hline
$CMB$&$PS$&$CMB+PS$&$CMB$ with holes&Residual&$CMB_{recon}$&Residual&$CMB_{recon}$   &Residual \\
     &    &        &                &  map   &             &map     &SNR$\approx 10$)&  map    \\
\hline
 4.10&4.75 & 6.23   &  3.97   & 0.99  &4.10   & 0   &  4.11     &  0.31   \\
\hline
\end{tabular}
\end{center}

\vskip 3cm

\begin{center}
\centerline{{\bf Table 2.}~Maps and their dispersions for
experiment group 2} \vskip 6mm
\begin{tabular}{|c|c|c|c|} \hline
Experiment& Map &Dispersion $\sigma$, $10^{-5}$&Dispersion of map with holes $\sigma$, $10^{-5}$\\
\hline
  2    &$CMB$              &4.10                   & 3.90 \\
       &$PS\ast BEAM$      &1.99                   &  0   \\
       &$CMB\ast BEAM$     &4.51                   & 4.29 \\
       &$(CMB+PS)\ast BEAM$&4.93                   & 4.29 \\
  2.1  & Pixelnoise        & 0                     & $-$  \\
       &$CMB_{meas}$       &4.93                   & $-$  \\
       &$CMB_{recon}$      &4.09                   & 3.89 \\
       &Residual map       &0.39                   & 1.28 \\
  2.2  &Pixelnoise         &0.65                   &  $-$ \\
       &$CMB_{meas}$       &4.97                   &  $-$ \\
       &$CMB_{recon}$      &4.05                   & 3.86 \\
       &Residual map       &0.74                   & 1.38 \\
  2.3  & Pixelnoise        &2.61                   &  $-$ \\
       &$CMB_{meas}$       &5.58                   &  $-$ \\
       &$CMB_{recon}$      &4.27                   & 4.10 \\
       &Residual map       &1.19                   & 1.59 \\
  2.4  &Pixelnoise         &5.23                   &  $-$ \\
       &$CMB_{meas}$       &7.19                   &  $-$ \\
       &$CMB_{recon}$      &4.14                   & 3.93 \\
       &Residual map       &1.47                   & 1.84 \\
\hline
\end{tabular}
\end{center}

\begin{figure}[d]
\centerline{\psfig{figure=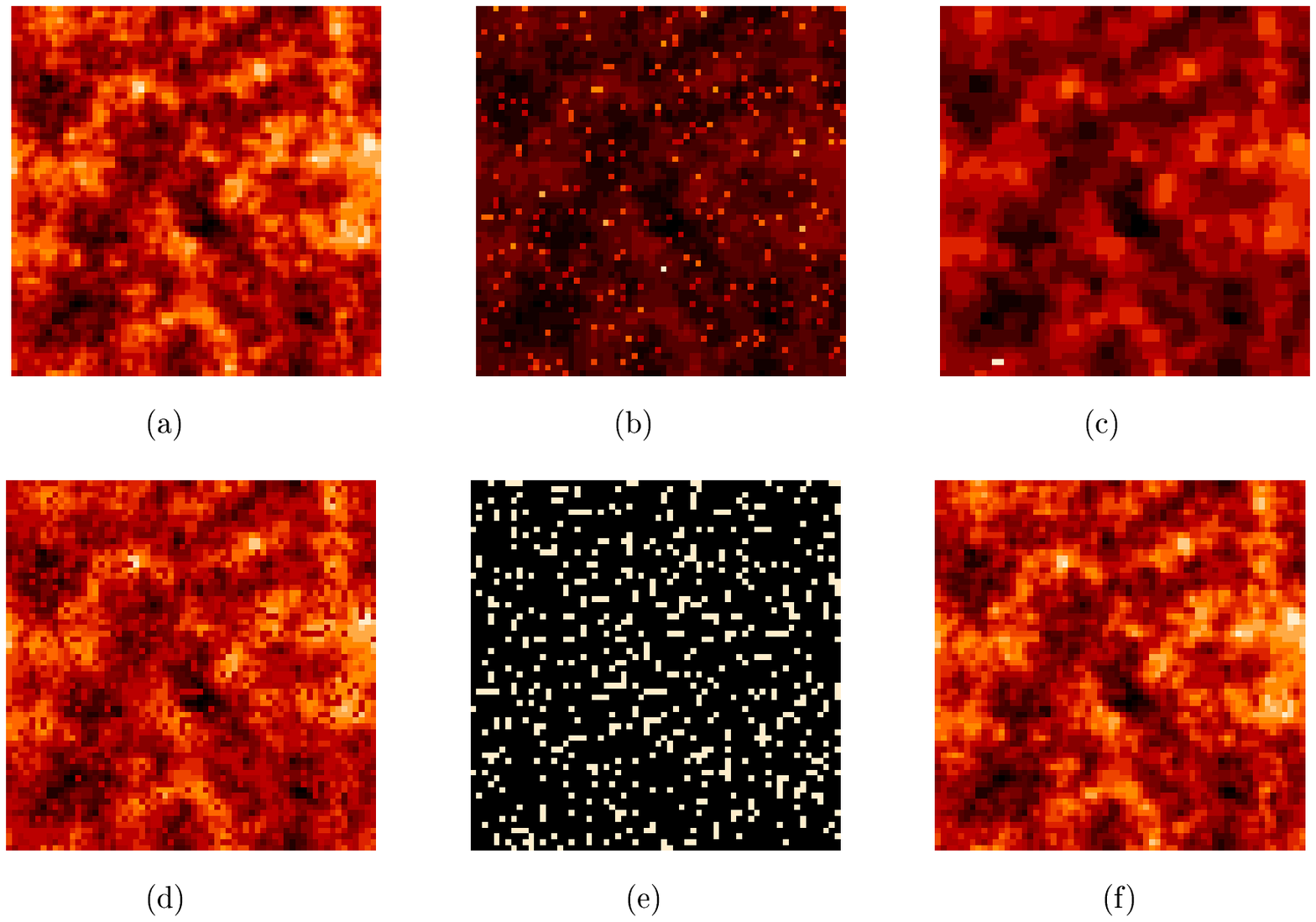,angle=0,width=150mm} }
\centerline{{\bf Fig.1.}~Maps for experiment 1.}
\end{figure}
\medskip

\begin{figure}[d]
\centerline{ \psfig{figure=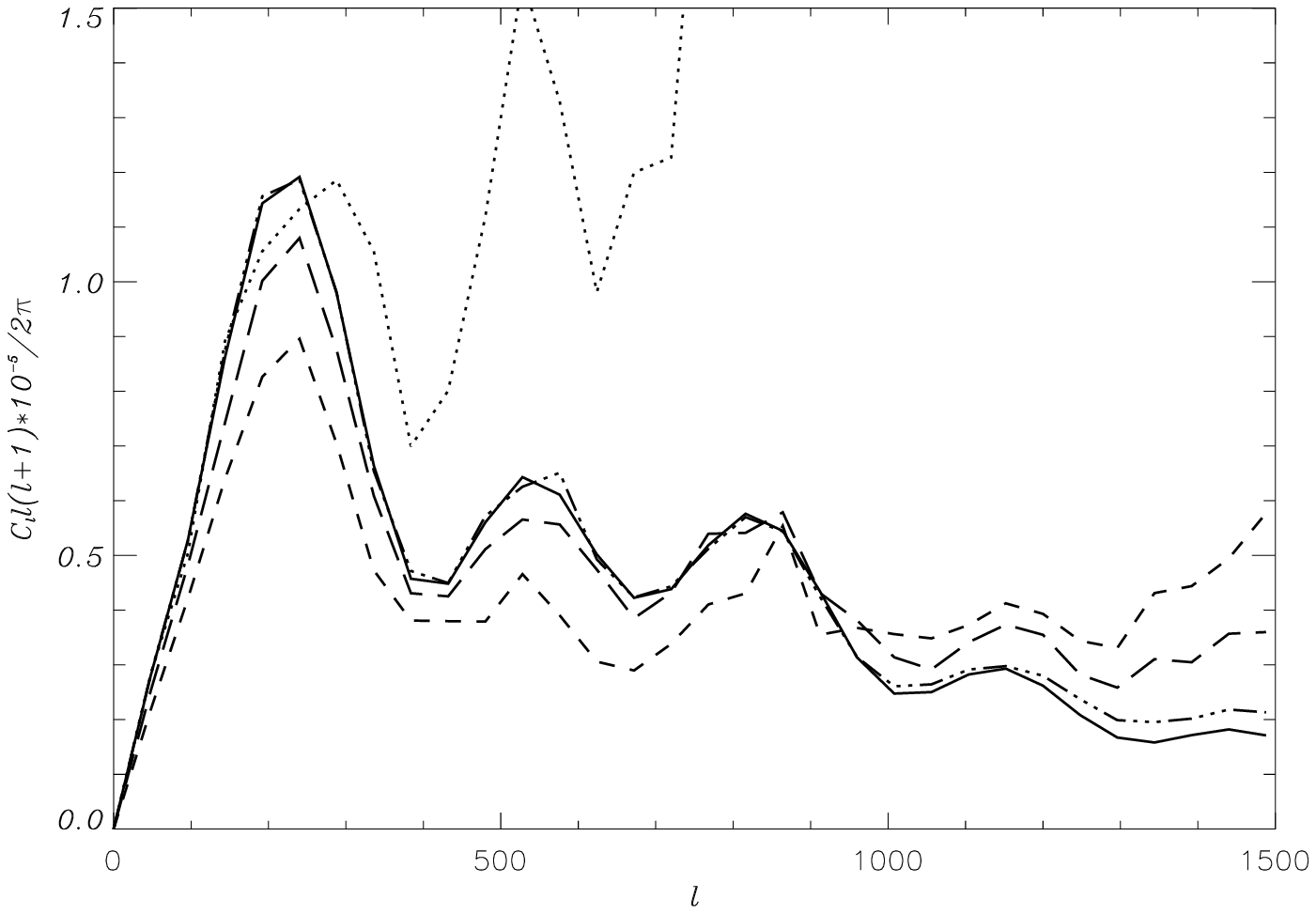,angle=0,width=150mm} }
\centerline{{\bf Fig.2.}~Angular power spectra for the maps for
experiment 1. See text for details.}
\end{figure}

\begin{figure}[d]
\centerline{ \psfig{figure=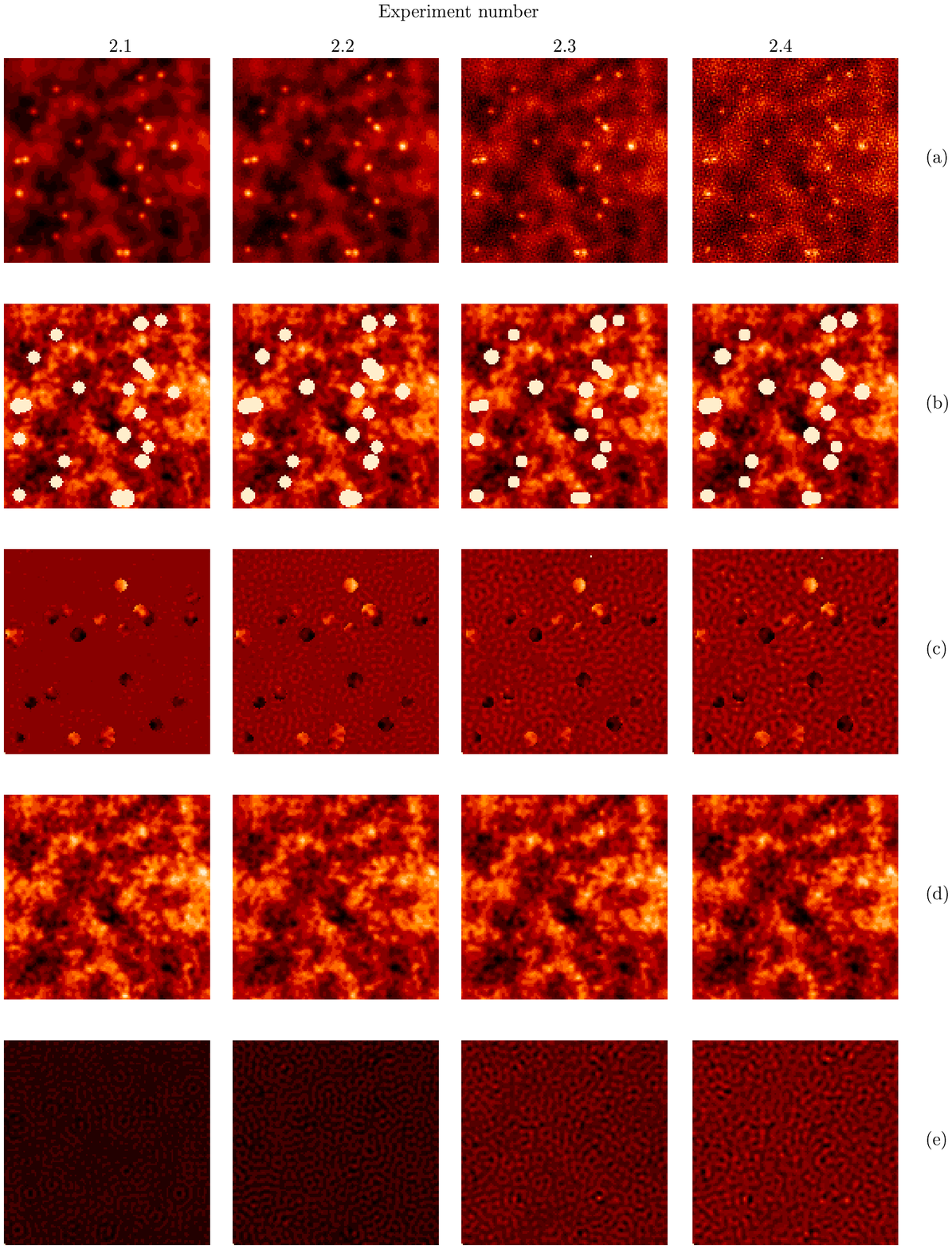,angle=0,width=150mm} }
\centerline{{\bf Fig.3.}~Maps for experiment group 2.}
\end{figure}

\begin{figure}[d]
\centerline{ \psfig{figure=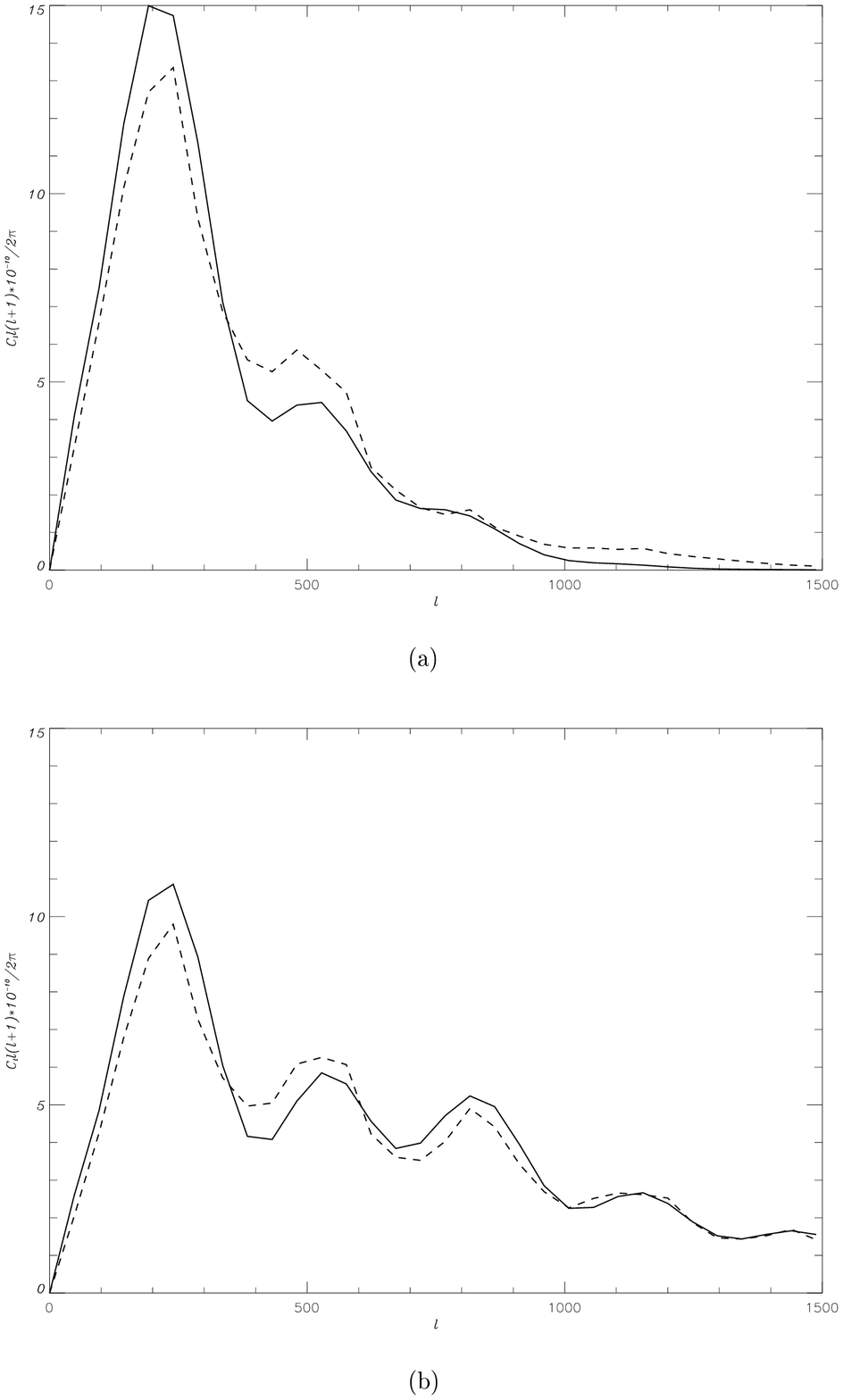,angle=0,width=130mm} }
\centerline{{\bf Fig.4.}~Illustrating the effect of excluding
sections contaminated by point sources. See text for details.}
\end{figure}
\medskip

\begin{figure}[d]
\centerline{ \psfig{figure=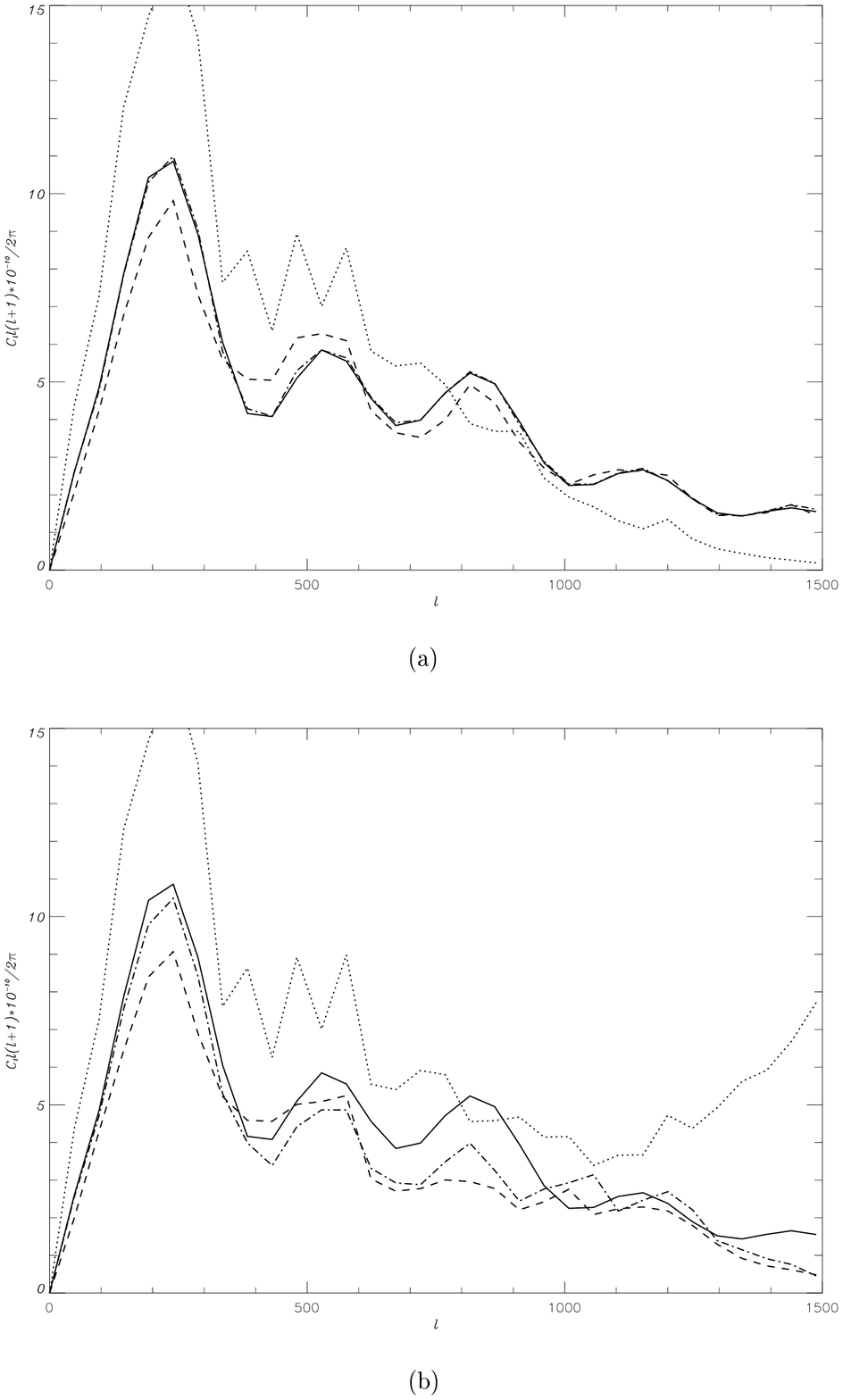,angle=0,width=130mm} }
\centerline{{\bf Fig.5.}~Angular power spectrum for the maps in
experiment group 2.See text for details.}
\end{figure}
\medskip

\begin{figure}[d]
\centerline{ \psfig{figure=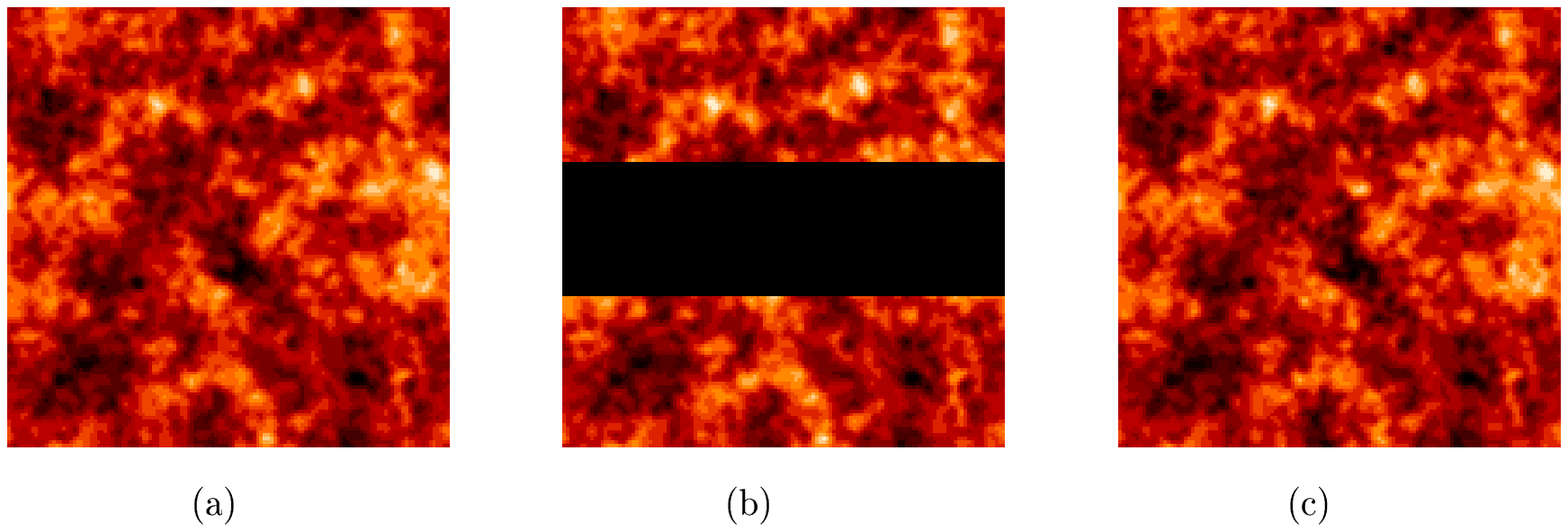,angle=0,width=150mm} }
\centerline{{\bf Fig.6.}~Maps for experiment 3.}
\end{figure}
\medskip

\begin{figure}[d]
\centerline{ \psfig{figure=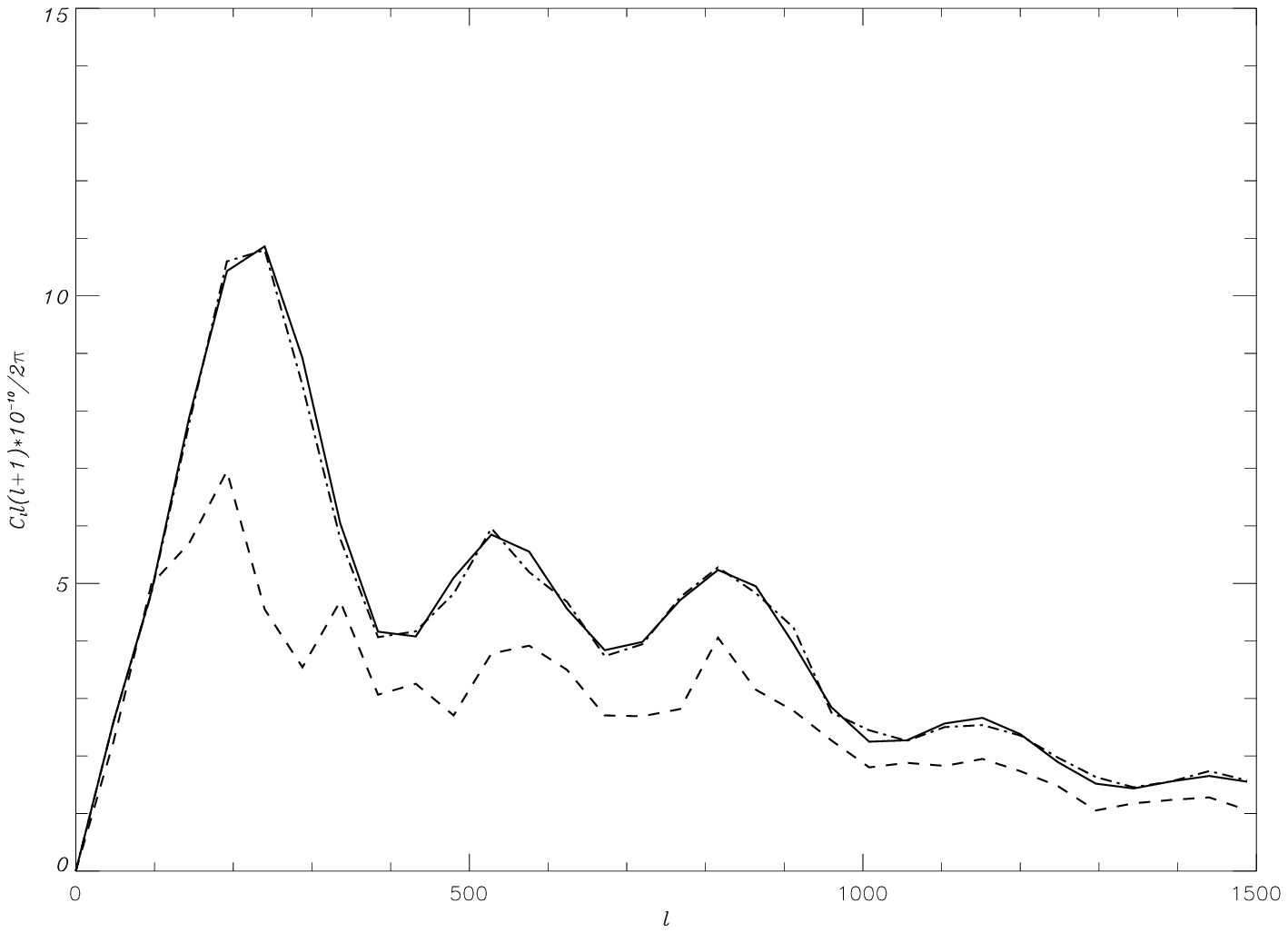,angle=0,width=150mm} }
\centerline{{\bf Fig.7.}~Angular power spectra for the maps for
experiment 3. See text for details.}
\end{figure}
\medskip

\end{document}